\documentclass[10pt,letterpaper]{article}

\usepackage[top=5cm, bottom=5cm, left=3.8cm, right=3.8cm]{geometry} 

\usepackage{mathptmx}
\usepackage{xcolor}

\usepackage{courier}

\usepackage{indentfirst}

\usepackage{booktabs}

\usepackage{graphicx}

\usepackage[font=small,labelfont=bf]{caption}

\usepackage{titling}
\pretitle{\begin{center}\fontsize{16pt}{19pt}\selectfont\bfseries}
\posttitle{\par\end{center}\vskip 1ex}

\renewenvironment{abstract}
{\begin{quote}
 \noindent {\small \bfseries \abstractname.} \small
}
{\medskip\noindent
 \end{quote}
}

\newenvironment{keywords}
{\begin{quote}
 \noindent {\small \bfseries Keywords: } \small
}
{\medskip\noindent 
 \end{quote}
}

\usepackage{titlesec}

\titleformat*{\section}{\large\bfseries}
\titleformat*{\subsection}{\normalsize\bfseries}
\titleformat*{\subsubsection}{\normalsize\bfseries}
\titleformat*{\paragraph}{\normalsize\bfseries}
\titleformat*{\subparagraph}{\normalsize\bfseries}

\usepackage[plain, noline]{algorithm2e}
\SetAlFnt{\fontfamily{pcr}\selectfont} 
\SetAlCapSkip{2ex}
\IncMargin{-1em}

\SetKwInput{KwIn}{Input}
\SetKwInput{KwOut}{Output}
\SetKwIF{If}{ElseIf}{Else}{if}{then}{else\ if}{else}{end\ if}
\SetKwFor{For}{for}{do}{end\ for}
\SetKw{KwTo}{until}
\SetKwRepeat{Repeat}{repeat}{until}
\SetKwFor{While}{while}{do}{end\ while}

\usepackage[hang,flushmargin]{footmisc} 

\usepackage{bibspacing}
\colorlet{myColor}{blue!50!black}

\makeatletter
\def\@eqnnum{{\normalfont \color{myColor} (\theequation)}}
\makeatother

\renewcommand\fbox{\fcolorbox{myColor}{white}}

\begin{document}

\title{5G-NR PRACH Detection Performance Optimization \\in Context of Intra/Inter-Cell Interference}

\author{
\normalsize Désiré~Guel$^{(1)}$, Pegdwindé~Justin~Kouraogo$^{(1)}$,  Boureima~Zerbo$^{(2)}$,  Elie~Jephte~Yaro$^{(1)}$  \\ 
\small $\lbrace$~guel.desire,~kouraogo,~jephteeyaro~$\rbrace$@gmail.com$^{(1)}$, $\lbrace$~bzerbo@gmail.com~$\rbrace$$^{(2)}$ 
\and 
\small Université Joseph KI-ZERBO (U-JKZ), Burkina Faso $^{(1)}$,  Université Thomas SANKARA (UTS), Burkina Faso $^{(2)}$
}

\date{}

\pagestyle{empty}

\maketitle
\thispagestyle{empty}

\vspace{-3.75em}

\begin{abstract}
The ever-evolving landscape of wireless communication technologies has led to the development of 5G-NR (5G New Radio) networks \cite{AMukherjee2019} promising higher data rates and lower latency. However, with these advancements come challenges in managing intra-cell and inter-cell interference, particularly during the random-access procedure. This article aims to explore the impact of UE (User Equipment) and cell configuration parameters on interference and establish improved interference management strategies. To assess the effectiveness of these strategies, Key Performance Indicators (KPIs) such as Correct Detection Rate (CDR) will be considered. Additionally, the Matched  Filtering (MF)-Based  PRACH (Physical Random-Access CHannel) detection algorithm \cite{Panasonic2007}  will be studied to evaluate the PRACH performance. 
The findings underscore the critical role of interference management and the configuration of User Equipment (UE)/Cell parameters in ensuring the robustness and reliability of the NR-PRACH performance in 5G NR networks. Optimizing these configurations can lead to improved network performance and enhanced communication quality.
\end{abstract}


\section{Introduction}

The advent of 5G NR (New Radio) technology has unlocked a new realm of possibilities in wireless communication, promising unprecedented speeds, ultra-low latency, and the capacity to connect billions of devices simultaneously \cite{AMukherjee2019}. However, the robustness and reliability of 5G NR networks in real-world scenarios depend heavily on the ability to effectively manage and mitigate interference, both within cells (intra-cell interference) and between neighboring cells (inter-cell interference).

In the pursuit of seamless connectivity and enhanced user experiences, the performance of the Random Access CHannel (PRACH) plays a key role. The 5G-NR PRACH serves as the gateway for User Equipment (UE) to initiate communication with the network, making it a critical component of the 5G NR architecture. However, its efficiency is profoundly influenced by the surrounding interference conditions.

Intra-cell interference arises when multiple UEs within the same cell attempt to access the network simultaneously, leading to contention for the PRACH resources. Conversely, inter-cell interference emerges when UEs from neighboring cells contend for the same PRACH resources, leading to contention across cell boundaries. These interference scenarios introduce complexities in PRACH detection, impacting network efficiency, latency, and overall Quality of Service (QoS).

This article embarks on a study to delve into the intricacies of 5G-NR PRACH detection performance optimization within the challenging backdrop of intra and inter-cell interference conditions. By comprehensively examining the impact of UE and cell configurations on PRACH performance. It  aims to unravel the key factors that influence detection reliability and devise strategies to enhance PRACH performance.


As 5G NR networks continue to evolve and expand their footprint, the insights gleaned from this research will not only contribute to the refinement of PRACH performance but also pave the way for more robust and interference-resilient 5G NR networks. In the pursuit of unlocking the full potential of 5G technology, the management of intra and inter-cell interference is a paramount endeavor, and this article endeavors to be a cornerstone in that endeavor.

\section{PRACH Principle and Preamble Detection}
This section presents the essential aspects of Random Access (RA)  in 5G-NR context. Preambles play a crucial role in synchronization, initial access, and signal detection. 

\subsection{Preamble  Structure and Procedure}
Random access (RA) is a basic procedure in 5G-NR technologies, enabling UEs to establish uplink synchronization and initiate uplink transmission. PRACH  employs phase modulation based on Zadoff‐Chu sequences \cite{Huawei2007} with distinct phase variations among different symbols within the sequences.  Before sending its random access request, UEs must retrieve a set of information transmitted by the base station through the SIB2 \textsl{(System Information Block Type 2)} message \cite{TS138331}. 

With this information, the UE can transmit the PRACH preamble using the  resources indicated by the Next-Generation NodeB (gNB) during the transmission of SIB2. The transmission of the PRACH is associated with the RA-RNTI \textsl{(Random Access RNTI)}.

\begin{figure}[htbp]
	\centering
\includegraphics[page = 1,clip, trim=0.0cm 0.0cm 0.0cm 0.0cm, width=0.95\textwidth]{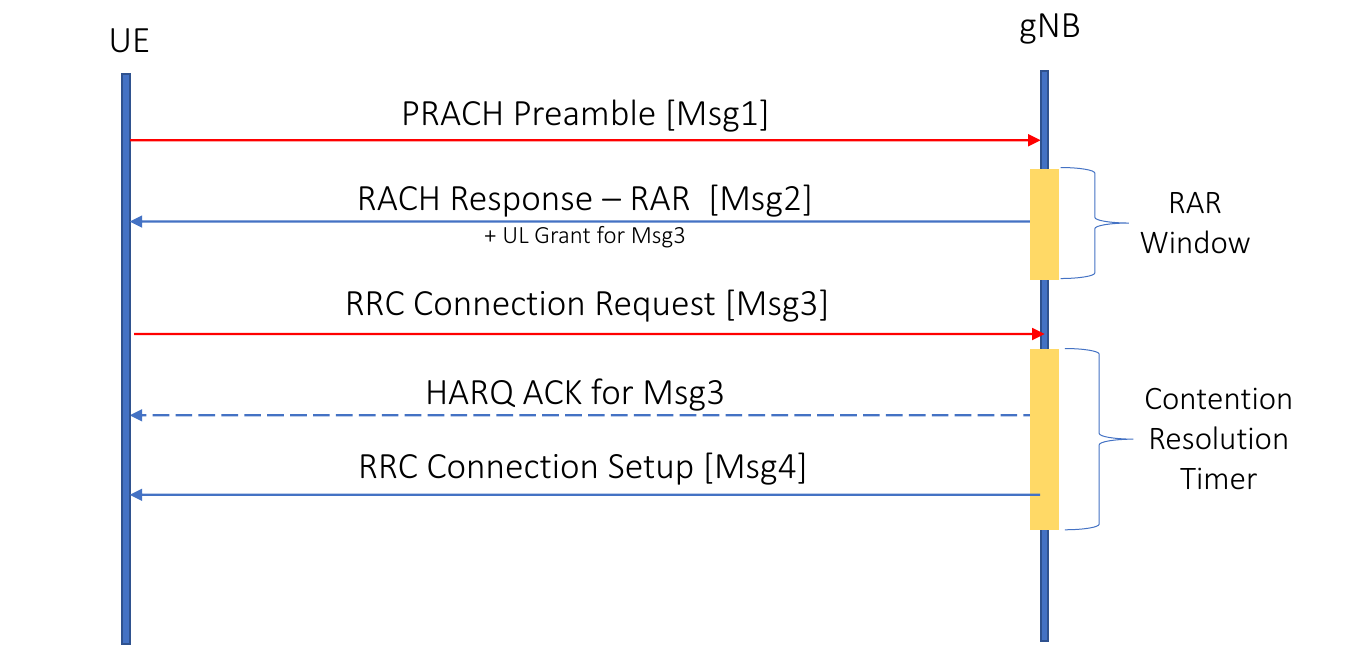}
	\caption{NR-PRACH Procedure.}
	\label{Fig:01}
\end{figure}

When an UE enters a new cell, it has no prior knowledge of the gNB. After identifying the optimal SSB \textsl{(Synchronization Signal  Block)}  through downlink synchronization, the UE transmits the PRACH containing its information, based on the best time index of the SSB. Fig.\ref{Fig:01} illustrates the interactions between the UE and the gNB during the initial access procedure \cite{AMukherjee2019}.

Random access process is described in Fig.\ref{Fig:01}. The UE selects randomly a preamble from a list of parameters broadcasted through the SIB2 and transmits it in the PRACH with an initial power result of a basic downlink pathloss estimation. If there is no answer from the gNB, the UE makes a retry with higher power level.

\begin{figure}[htbp]
	\centering
\includegraphics[page = 1,clip, trim=0.0cm 0.0cm 0.0cm 0.0cm, width=0.65\textwidth]{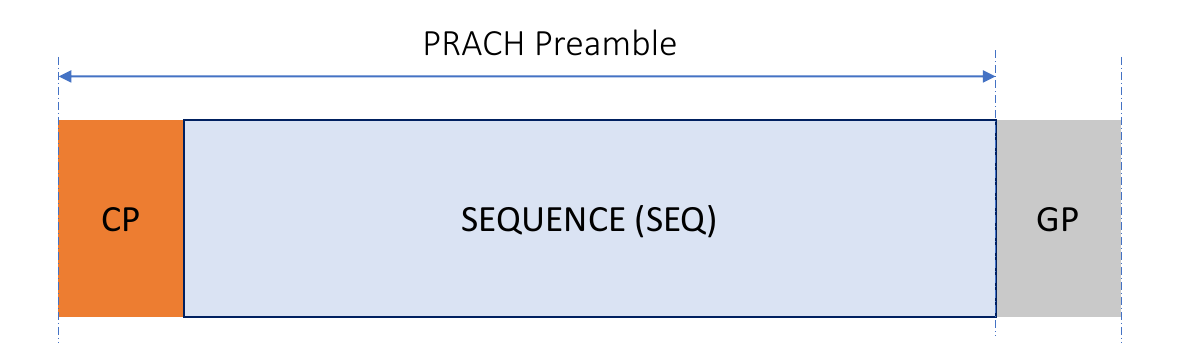}
	\caption{NR-PRACH structure.}
	\label{Fig:02}
\end{figure}

The 5G-NR PRACH preamble consisted of a complex sequence (SEQ) which an OFDM symbol, built with a Cyclic Prefix (CP), thus allowing for an efficient frequency-domain receiver at the gNB. The preamble length is shorter than the PRACH slot in order to provide room for a Guard Period (GP) to absorb the propagation delay. The CP facilitates PRACH processing in the frequency domain. Refer to Fig.\ref{Fig:02} for the NR-PRACH structure.

\subsection{Preamble Sequence Generation}
The preamble sequence length is set to a prime number of 839, there are 838 sequences with optimal cross‐correlation properties. The ${{\rm{u}}^{{\rm{th}}}}$ $\left({{\rm{0}} \le {\rm{u}} \le {\rm{837}}} \right)$ root Zadoff‐Chu sequence is defined by (Nzc is the length of the Zadoff‐Chu sequence):
\textcolor{myColor}{
\begin{equation}
{x_u}\left( n \right) = {e^{ - j\frac{{\pi un\left( {n + 1} \right)}}{{{N_{ZC}}}}}},0 \le n \le {N_{ZC}} - 1
\label{Eq:01}
\end{equation} 
}


From the ${{\rm{u}}^{{\rm{th}}}}$root ZC sequence, random access preambles with Zero Correlation Zones \textsl{(ZCZ)} of length ${N_{ZC}} - 1$ are defined by Cyclic Shifts \textsl{(CS)} according to \cite{3GPPTS36211V890}:

\textcolor{myColor}{
\begin{equation}
{x_{u,v}}\left( n \right) = {x_u}\left( {\left( {n + {C_v}} \right)\bmod {N_{ZC}}} \right),
\label{Eq:02}
\end{equation}  
}
where ${{C_v}}$ is the cyclic shift, and ${{N_{ZC}}}$ is the cyclic shift offset. This paper adopts preamble format 0 in the 5G-NR, which generates from a 839 point ZC sequence which is  specifically designed for contention-based access, where multiple UEs may attempt to access the network simultaneously. 
 

\subsection{Preamble Detection }
The PRACH receiver is implemented to maximize the probability of ``correct'' preamble detection and minimize processing latency. The functional diagram of the conventional PRACH receiver, which shares certain operations with the Orthogonal Frequency Division Multiplexing (OFDM) demodulator, is illustrated in Fig.\ref{Fig:03}. The 5G-NR PRACH receiver process \cite{SSesia2011} :
\begin{description}
	\item [(1)] \textcolor{myColor}{\textsl{ CP-GP (Cyclic Prefix and Guard Period) Removal}}: In this step, the Cyclic Prefix (CP) and Guard Period (GP), which are added to the transmitted PRACH signal for synchronization purposes, are removed to obtain the original signal.
	\item [(2)] \textcolor{myColor}{\textsl{Frequency Shift}}: The received signal is shifted in frequency to bring it to a common reference frequency, aligning it with the system's reference.
	\item [(3)] \textcolor{myColor}{\textsl{Decimation}}: The signal is sampled at a lower rate, reducing the data rate while retaining essential information, to simplify subsequent processing.
	\item [(4)] \textcolor{myColor}{\textsl{FFT (Fast Fourier Transform)}}: An FFT is applied to the decimated signal to transform it from the time domain to the frequency domain, allowing the identification of frequency components.
	\item [(5)] \textcolor{myColor}{\textsl{Sub-Carrier Demapping}}: In this step, sub-carriers used for the PRACH signal are demapped, extracting the relevant information from the frequency domain representation.
	\item [(6)] \textcolor{myColor}{\textsl{IFFT (Inverse Fast Fourier Transform)}}: An IFFT is performed to convert the signal back from the frequency domain to the time domain, preparing it for further processing.
	\item [(7)] \textcolor{myColor}{\textsl{Signature Detection}}: Finally, signature detection technique is applied to identify and extract specific patterns or signatures in the PRACH signal, which are crucial for synchronization and channel estimation in 5G-NR communication systems. Fig.\ref{Fig:04a} shows the basic functions of the signature detector.
\end{description}


\begin{figure}[htbp]
	\centering
\includegraphics[page = 1,clip, trim=0.0cm 0.0cm 0.0cm 0.0cm, width=1.05\textwidth]{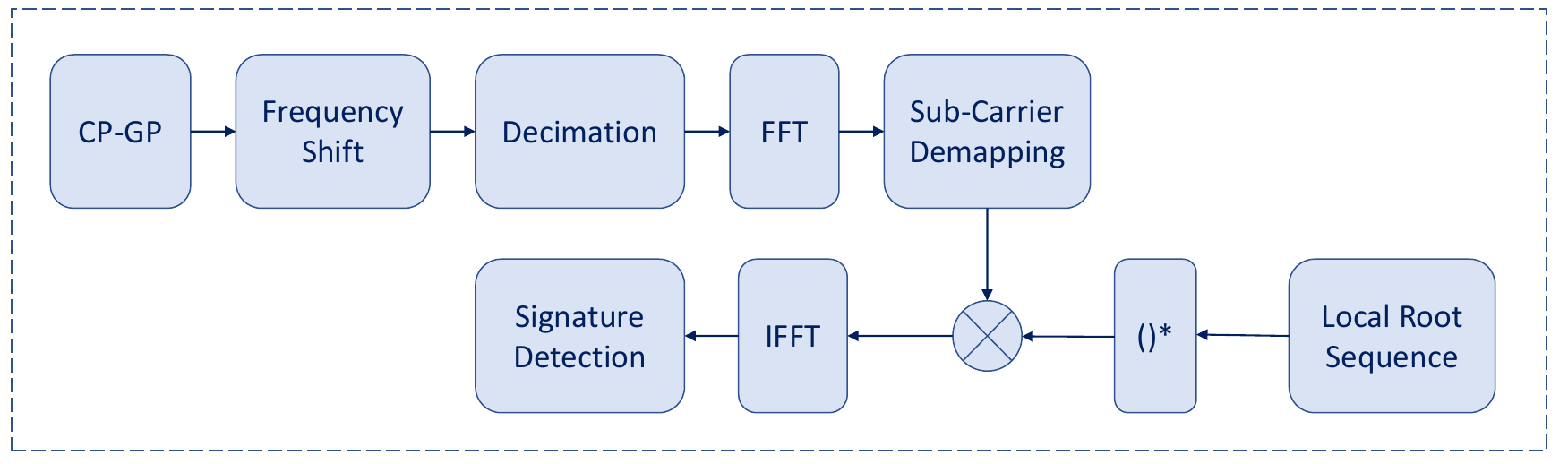}
	\caption{5G-NR PRACH Receiver Block Diagram.}
	\label{Fig:03}
\end{figure}


The fact that different PRACH signatures are generated from Cyclic Shifts \textsl{(CS)} of a common root sequence means that the frequency-domain computation of the Power Delay Profile \textsl{(PDP)} of a root sequence provides in one shot the concatenated PDPs of all signatures derived from the same root sequence. Therefore, the signature detection process consists of searching, within each ZCZ defined by each Cyclic Shifts (CS), the PDP peaks above a detection threshold over a search window corresponding to the cell size. 

\begin{figure}[htbp]
	\centering
\includegraphics[page = 1,clip, trim=0.0cm 0.0cm 0.0cm 0.0cm, width=1.05\textwidth]{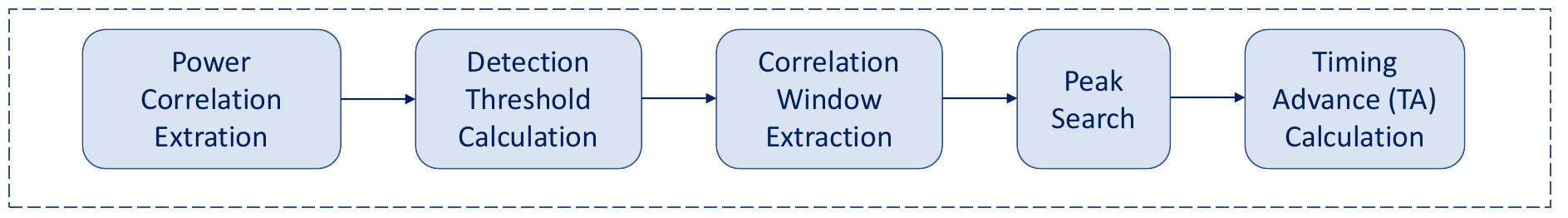}
	\caption{Preamble Detection Block Diagram}
	\label{Fig:04a}
\end{figure}

%

\begin{description}
	\item [(1)]  \textsl{\textcolor{myColor}{Power Correlation Extraction}}: In this step, the received signal's power is correlated with a known ``root sequence'' to identify potential preamble candidates within the received signal. This process helps in locating the start of the frame.
	\item [(2)]  \textsl{\textcolor{myColor}{Detection threshold Calculation}}. The target false alarm probability ${p_{fa}}\left( {{T_{\det }}} \right)$ drives the setting of the detection threshold ${{T_{\det }}}$. In \cite{SSesia2011}, it is showed that under the assumption that the $L$ samples in the uncertainty window are uncorrelated Gaussian noise with variance $\sigma _n^2$ in the absence of preamble transmission, the complex sample sequence $z_a^m\left( \tau  \right)$ received from antenna a (delayed to reflect a targeted time offset $\tau$ of the search window, and despread over a coherent accumulation length (in samples) ${N_{ca}}$ against the reference code sequence) is a complex Gaussian random variable with variance $\sigma _{n,ca}^2 = {N_{ca}} \times \sigma _n^2$. In practice, ${N_{ca}}$ is the size of the IFFT. The non-coherent accumulation ${z_{nca}}\left( \tau  \right)$ is modelled as follows \cite{SSesia2011} :
\textcolor{myColor}{
\begin{equation}
{z_{nca}}\left( \tau  \right) = \sum\limits_{a = 1}^{{N_a}} {\sum\limits_{m = 0}^{{N_{nca}} - 1} {{{\left| {z_a^m\left( \tau  \right)} \right|}^2}} } 
\label{Eq:03}
\end{equation}
}
where ${{N_a}}$ is the number of antennas and ${{N_{nca}}}$ is the number of additional non-coherent accumulations (e.g. in case of sequence repetition).

${z_{nca}}\left( \tau  \right)$ follows a central chi-square distribution with
$2N = 2{N_a} \times {N_{nca}}$ degrees of freedom, with mean (defining the noise floor) ${\gamma _n} = N \times \sigma _{n,ca}^2$ and Cumulative Distribution Function (CDF) $F\left( {{T_{\det }}} \right) = 1 - {p_{fa}}{\left( {{T_{\det }}} \right)^L}$. It  is worth noticing that instead of the absolute threshold , in \cite{SSesia2011} the authors consider the threshold ${T_r}$ relative to the noise floor ${\gamma _n}$ as follows:
\textcolor{myColor}{
\begin{equation}
{T_r} = \frac{{{T_{\det }}}}{{{\gamma _n}}} = \frac{{{T_{\det }}}}{{{N_a} \times {N_{nca}}{N_{ca}}\sigma _n^2}}
\label{Eq:04}
\end{equation}
}
	\item [(3)] \textsl{\textcolor{myColor}{Correlation Window Extraction:}} The received signal is divided into smaller time intervals or windows to focus the correlation process on specific parts of the signal. These windows allow for more precise identification of the preamble within the signal.
	\item [(4)] \textsl{\textcolor{myColor}{Peak Search:}} Within each correlation window, the highest correlation peak is identified. This peak corresponds to the position where the preamble sequence aligns most closely with the received signal, indicating the start of the frame.
  \item [(5)] \textsl{\textcolor{myColor}{Timing Advance (TA) Calculation:}} Once the peak is found, the timing advance (TA) is calculated to determine the timing misalignment between the received signal and the expected frame timing. The TA is used to synchronize the receiver with the incoming signal for further data reception and processing.
\end{description}

\section{Related Works}


This section explores various studies \cite{Q.Xiong2018,GSchreiber2018,ZGuo2020} related to the performance of the Physical Random Access Channel (PRACH). However we can highlight the scarcity of research in this area of 5G-NR PRACH performance in conditions of intra-cell and inter-cell interference, underscoring the importance of the originality and relevance of the present study. We also recognizes the existence of relevant prior work related to PRACH performance in different contexts. This section provides a valuable foundation for understanding the context and significance of the current study's contribution.  In the following subsections, we review and discuss some of these noteworthy studies.

\subsection{Random Access Preamble Generation and Procedure Design for 5G-NR System \cite{Q.Xiong2018}}
This article  addresses challenges posed by the transition from LTE to 5G-NR systems, focusing on Physical Random Access Channel (PRACH) preamble generation and random access protocols. It proposes an OFDM-based signal generation technique that aims to mitigate potential interference from neighboring subcarriers.
\begin{description}
	\item [(a)] \textsl{\textcolor{myColor}{Strengths:}}
	\begin{itemize}
		\item The proposed solution in \cite{Q.Xiong2018} accounts for the fundamental differences between LTE and 5G-NR systems, especially regarding multi-subcarrier spacing (SCS) and multi-analog-beam-powered operations.
		\item It adopts an OFDM-based approach, which is well-suited for 5G-NR systems, offering advantages in terms of spectral efficiency and flexibility.
		\item The consideration of multi-beam capability is a significant strength, as it aligns with the beamforming and massive MIMO features of 5G-NR networks.
\end{itemize}
 \item [(b)] \textsl{\textcolor{myColor}{Weaknesses:}}
 \begin{itemize}
	 \item  While the article mentions the consideration of potential interference from neighboring subcarriers, it lacks a detailed analysis or simulation results to demonstrate the robustness and reliability of the proposed PRACH preamble generation method in real interference conditions.
	 \item  The proposed solution in \cite{Q.Xiong2018}  doesn't explicitly discuss the measures taken to ensure the reliability of the proposed random access protocol, especially in scenarios with high user density or network congestion.
\end{itemize}
 \item [(c)] \textsl{\textcolor{myColor}{Relevance  with the current study:}}
The proposed solution in \cite{Q.Xiong2018} is relevant to our study.  While \cite{Q.Xiong2018} focuses on the generation and protocol design of PRACH preambles, the relevance lies in the fact that effective interference management is crucial for successful PRACH detection, especially in interference-prone scenarios.
\end{description}

\subsection{5G New Radio Physical Random Access Preamble Design \cite{GSchreiber2018}}

The article proposes a RA preamble design based on cyclically delay-Doppler shifted m-sequences, which is presented as more robust against frequency uncertainties caused by wireless channel propagation and local oscillator imperfections compared to the legacy 4G LTE RA preambles based on Zadoff-Chu (ZC) sequences.

\begin{description}
	\item [(a)] \textsl{\textcolor{myColor}{Strengths:}}
\begin{itemize}
	\item The use of m-sequences is claimed to enhance robustness, making the proposed design more reliable in 5G NR networks.
	\item Simulation results are mentioned to support the claim that m-sequence based preambles perform well even under harsh transmission conditions.
\end{itemize}	
	\item [(b)] \textsl{\textcolor{myColor}{Weaknesses:}}
\begin{itemize}
	\item  While the article discusses the advantages of m-sequence based preambles, it does not thoroughly analyze potential limitations or trade-offs associated with this approach.
\end{itemize}	
	\item [(c)] \textsl{\textcolor{myColor}{Ability to manage and mitigate interference:}}
\begin{itemize}
	\item  It focuses primarily on the robustness of the RA preamble design against frequency uncertainties and local oscillator imperfections.
	\item  It highlights advantages such as RACH capacity enhancement, support for low-power devices, and low-complexity implementation.
	\item  It lacks information regarding interference management and mitigation strategies, which are critical aspects in the context of 5G NR networks.
	\item  Interference management is a crucial factor in the performance of wireless networks, and it does not provide insights into how the proposed design addresses interference issues.
\end{itemize}	
\end{description}

\subsection{5G NR Uplink Coverage Enhancement Based on DMRS Bundling and Multi-slot Transmission \cite{ZGuo2020}}

The article focuses on enhancing the uplink coverage in 5G NR networks, particularly addressing the limitation of coverage at cell edges. It proposes the use of demodulation reference signals (DMRS) bundling and multi-slot transmission to improve the performance of physical uplink shared channel (PUSCH) coverage.

\begin{description}
	\item [(a)] \textsl{\textcolor{myColor}{Strengths:}}
\begin{itemize}
	\item  The proposed solution introduces the bundling of DMRS in the same slot and repetitive transmission of the same data in multiple slots, which enhances the accuracy of channel estimation and improves coverage.
	\item  Simulation results demonstrate the effectiveness of the proposed schemes in improving coverage.
\end{itemize}	
	\item [(b)] \textsl{\textcolor{myColor}{Weaknesses:}}
\begin{itemize}
	\item While the proposed solution addresses coverage limitations, it doesn't extensively discuss other potential issues or interference sources that can affect 5G NR network reliability.
\end{itemize}	
	\item [(c)] \textsl{\textcolor{myColor}{Relevance with the current study:}}
\begin{itemize}
	\item  The  article \cite{ZGuo2020} focuses on coverage enhancement, while the current study focuses on PRACH detection performance in interference conditions.
	\item  There is indirect relevance since improved coverage can potentially contribute to better overall network performance, including PRACH detection.
\end{itemize}	
\end{description}

\section{Research Methodology}

The research methodology used in this study, consisted of evaluating and analyzing  5G-NR PRACH  performance in the context of intra-cell and inter-cell interference using the appropriate simulation tool Matlab \cite{MathWorks2022}. The simulation framework and setup parameters are presented in detail to provide insights into the conducted experiments and assessments. Additionally, we discuss the considerations for both intra-cell and inter-cell interference, which play a pivotal role in the accurate representation of real-world scenarios.

\subsection{Simulation Setup Parameters} 


To conduct our investigations, we harnessed the capabilities of the Matlab tool \cite{MathWorks2022}, which inherently integrates functionalities of the 5G-NR transmission chain. This powerful tool enabled us to simulate various aspects of the PRACH efficiently. Table \ref{Tab:01}, Table \ref{Tab:02} and Table \ref{Tab:03} serve as  references, enumerating the parameters that guided our simulation process. 

\begin{table}[htbp]
\textcolor{myColor}{
\centering
\caption{User Equipment (UE) Configuration}
\label{Tab:01}
\resizebox{0.99\textwidth}{!}{
\begin{tabular}{lccp{6cm}}
\toprule
               & ``Target'' UE   & ``Interfering'' UE  &                                     \\ 
               & (T-UE)  & (I-UE) &                                     \\ 
               &  Configuration       & Configuration     & Description                         \\ 
\midrule
NULRB          &  6                   & 6                 & Number of  UL Resource Blocks (RB)  \\ 
DuplexMode     &  FDD                 & FDD               & Frequency Division Duplexing (FDD)  \\ 
CyclicPrefixUL &  Normal              & Normal            & Normal cyclic prefix (CP)           \\ 
NTxAnts        &  1                   & 1                 & Number of transmit antennas         \\ 
\bottomrule
\end{tabular}
}
}
\end{table}

\begin{table}[htbp]
\textcolor{myColor}{
\centering
\caption{PRACH Configuration}
\label{Tab:02}
\resizebox{0.99\textwidth}{!}{
\begin{tabular}{lccl}
\toprule
               &  PRACH Configuration       & PRACH Configuration     &                          \\ 
               &  for ``Target'' UE  & for  ``Interfering'' UE  &                                     \\ 
               &   (T-UE) &  (I-UE) &             Description                        \\ 
\midrule
Format          &  0   & 0    & PRACH format:         \\
          &     &     & TS36.104, Table 8.4.2.1-1         \\
SeqIdx          &  22  & [0, 1, 2, 3, 4, 22]  & Root Sequence Index                          \\
CyclicShiftIdx  &  1   & 1    & Cyclic shift index                              \\
HighSpeed       &  0   & 0    & Normal mode, i.e. \\
       &     &     &  not optimized for ``HighSpeed'' \\
FreqOffset      &  0   & 0    & Default frequency location                      \\
PreambleIdx     &  32  & [0, 3, 37, 42, 63] & Preamble Index                                  \\
\bottomrule
\end{tabular}
}
}
\end{table}

\begin{table}[htbp]
\textcolor{myColor}{
\centering
\caption{Propagation channel configuration}
\label{Tab:03}
\resizebox{0.99\textwidth}{!}{
\begin{tabular}{p{3cm}cp{6cm}}
\toprule
               &  Channel Configuration       & Description                   \\ 
\midrule
NRxAnts             & 2         &  Number of receiving antennas               \\
DelayProfile        & `` ETU''   &  Delay profile                              \\
DopplerFreq         & 70.0 [Hz]      &  Doppler frequency                          \\
MIMOCorrelation     & `` Low ''   &  MIMO correlation                           \\
Seed                & 1         &  Channel seed                               \\
NTerms              & 16        &  Oscillators used in the fading pattern     \\
ModelType           & `` GMEDS '' &  Rayleigh fading model type                 \\
InitPhase           & `` Random '' & Phases initiales aléatoires                \\
NormalizePathGains  & `` On ''    &  Normalize delay profile power              \\
NormalizeTxAnts     & `` On ''   &  Normalize for transmit antennas            \\
\bottomrule
\end{tabular}
}
}
\end{table}


\subsection{Intra-Cell and Inter-Cell Interference Considerations}

In the realm of 5G-NR communication, the presence of intra-cell and inter-cell interference is a critical aspect that demands meticulous consideration. We meticulously modeled these interference within our Matlab simulator \cite{MathWorks2022} to ensure the accuracy and reliability of our simulation outcomes.

\begin{figure}[htbp]
	\centering
\includegraphics[page = 1,clip, trim=0.0cm 0.0cm 0.0cm 0.0cm, width=0.75\textwidth]{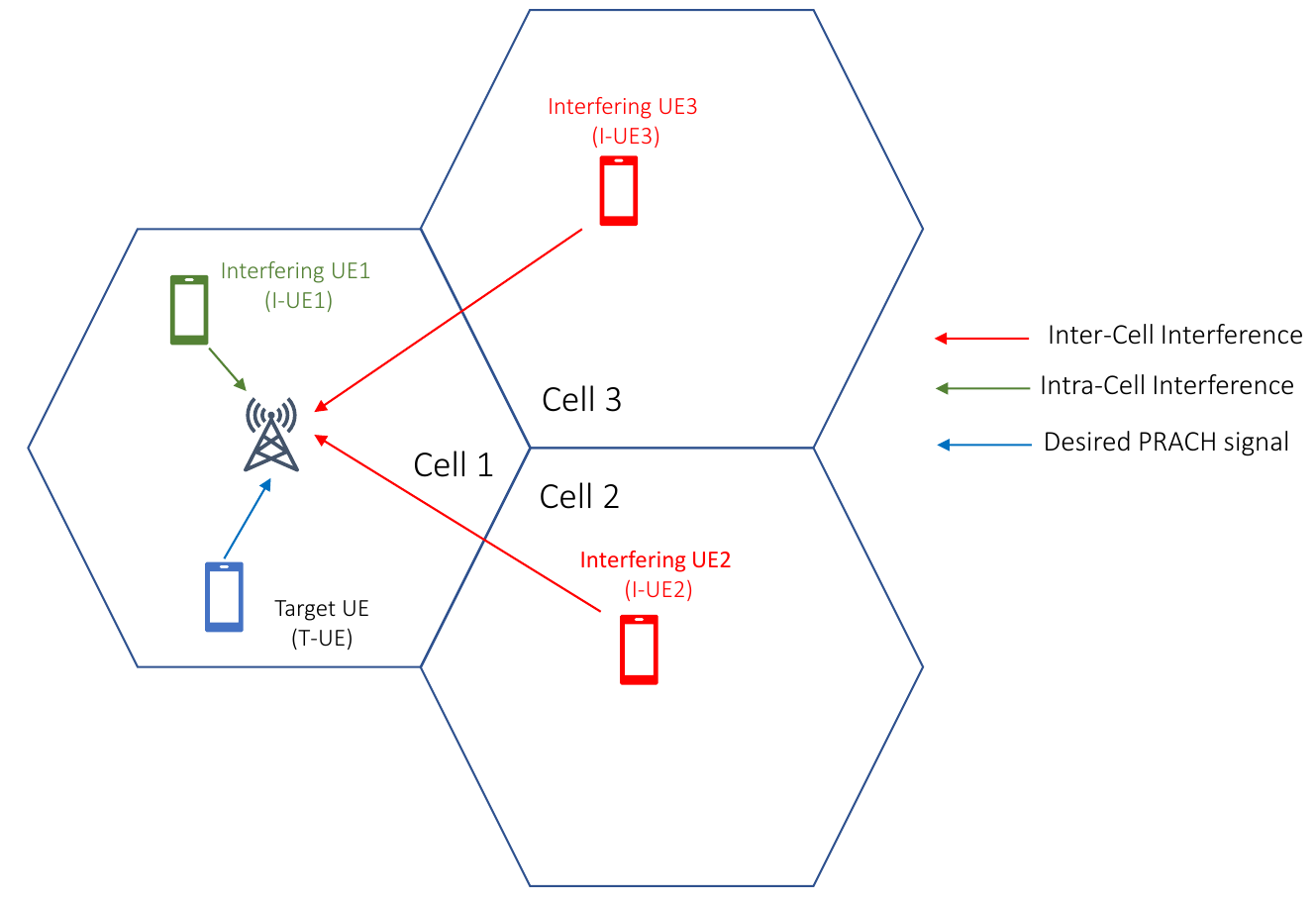}
	\caption{Inter-Cell and Intra-Cell Interference in 5G-NR PRACH Scenario.}
	\label{Fig:03}
\end{figure}

\begin{description}
	\item [(1)] \textsl{\textcolor{myColor}{Intra-Cell Interference}}: In the Matlab simulator \cite{MathWorks2022}, intra-cell NR-PRACH interference are modeled using an ``Interfering'' UE (i.e., the UE generating the interference) with the same UE configuration (refer to Table \ref{Tab:01}) and the same PRACH configuration (refer to Table \ref{Tab:01}), except for the ``Preamble Index'', which differs from whose of the ``Target UE'' (T-UE). This scenario is exemplified by ``Interfering'' ${\rm{UE}}_{\rm{1}}$ (${\rm{I-UE}}_{\rm{1}}$ ) in Fig. \ref{Fig:03}. 
	\item [(2)] \textsl{\textcolor{myColor}{Inter-Cell Interference}}: In the case of NR-PRACH inter-cell interference, it is modeled using a  ``Interfering''  UE   that shares the same UE configuration (refer to Table \ref{Tab:01}) and the same PRACH configuration (refer to Table \ref{Tab:02}) with the ``Target'' UE with the exception of the ``Logical sequence index'' or ``root sequence index'', which differs from whose of the ``target'' UE. It corresponds to the ``Interfering'' ${\rm{UE}}_{\rm{2}}$ and  ``Interfering'' ${\rm{UE}}_{\rm{3}}$  of Fig.\ref{Fig:03}. In fact, in a 5G-NR cell, all UEs within the cell share the same ``Logical Sequence Index'' or ``Root Sequence Index,'' while the UEs themselves are differentiated from each other using a ``Preamble Index''. The parameter ``Logical Sequence Index'' or ``Root Sequence Index'' serves to distinguish 5G-NR cells from one another. For example ``Target'' UE (T-UE) and  ``Interfering'' ${\rm{UE}}_{\rm{1}}$  of Fig.\ref{Fig:03} are sharing the  same ``Logical Sequence Index'' or ``Root Sequence Index but they  have different ``Preamble Index'' for Random-Access (RA).
\end{description}



\section{Results and Discussion}
\label{sec:sss}

In this section, we present the simulation results obtained using MATLAB by modeling an uplink transmission chain corresponding to the initial access procedure, which occurs through the Physical Random Access Channel (PRACH). The diagrams in Fig.\ref{Fig:01} and  Fig.\ref{Fig:02} depicts the simulated model in two scenarios: (i) An initial access context with the presence of intra-cell interference and (ii) An initial access context with the presence of inter-cell interference as illustrated in Fig.\ref{Fig:03}.

\subsection{NR-PRACH Performance under Intra-Cell Interference Conditions}
\label{sec:sss1}
The performance of NR-PRACH is assessed through the Correct Detection Rate (CDR) or Preamble Detection Probability carried out over 1000 subframes. Intra-cell interference are simulated using a ``Interfering'' UE  (I-UE) located in the same cell as the ``Target'' UE (T-UE) . This implies they share the same ``Logical Sequence Index'' but employ a different ``Preamble Index'' compared to T-UE.

\begin{figure}[htbp]
	\centering
	\fbox{
\includegraphics[page = 1,clip, trim=0.0cm 0.0cm 0.0cm 0.0cm, width=0.99\textwidth]{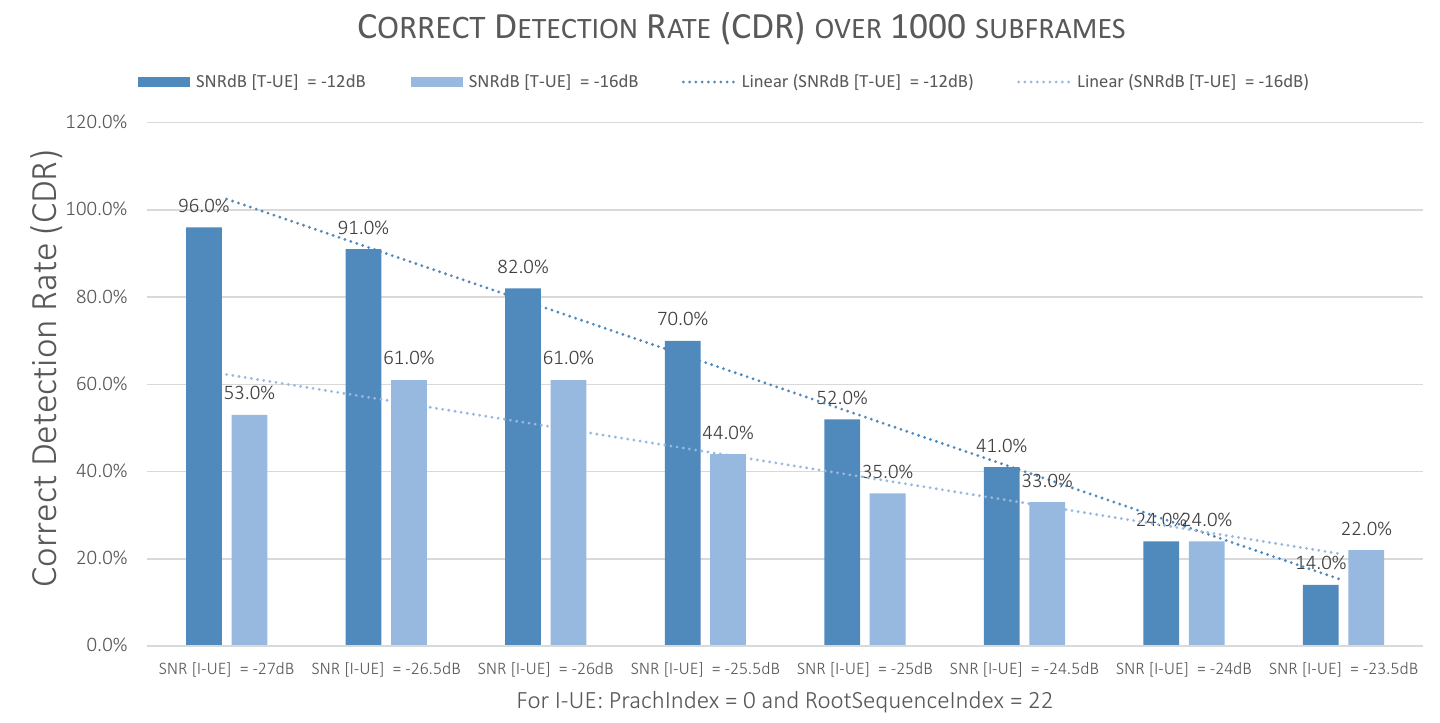}
}
	\caption{5G-NR PRACH Performance in the Presence of Intra-Cell Interference with Preamble Index = 0 and Root Sequence Index = 22}
	\label{Fig:04}
\end{figure}

\begin{figure}[htbp]
	\centering
	\fbox{
\includegraphics[page = 1,clip, trim=0.0cm 0.0cm 0.0cm 0.0cm, width=0.99\textwidth]{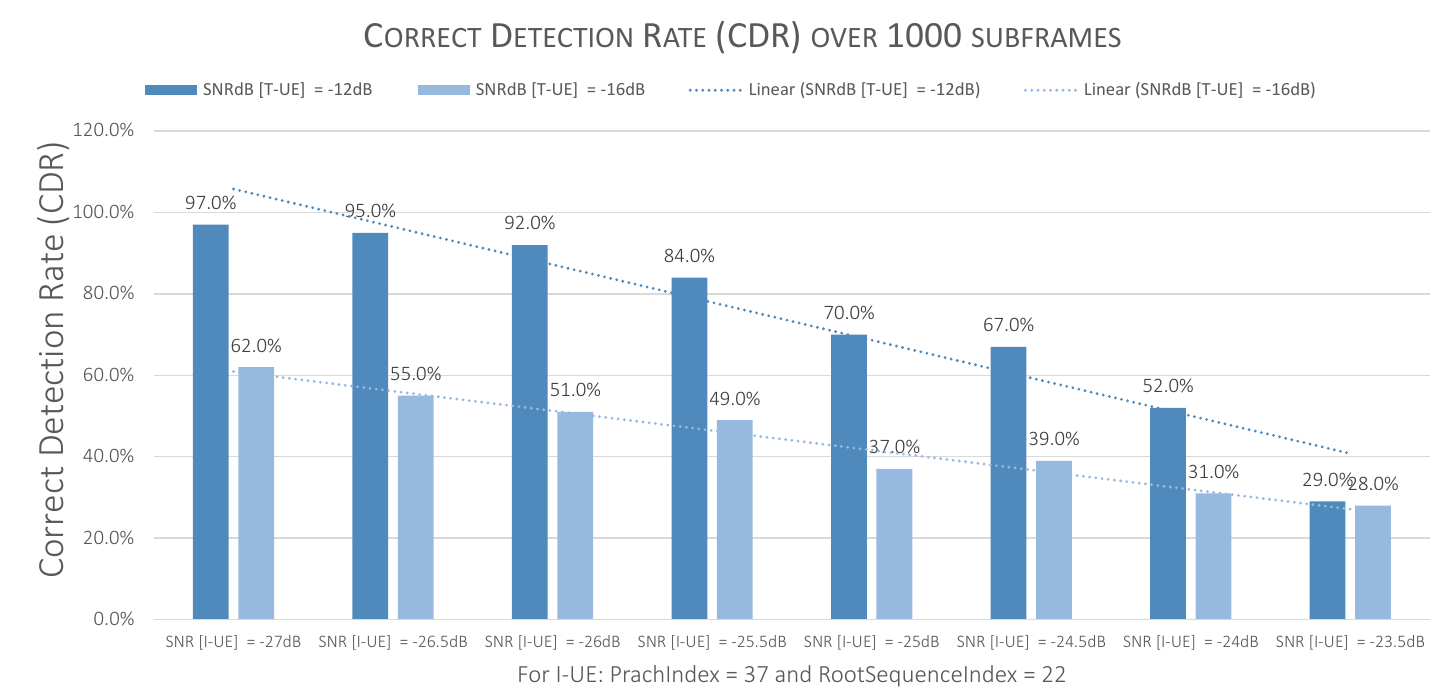}
}
	\caption{5G-NR PRACH Performance in the Presence of intra-cell interference with Preamble Index = 37 and Root Sequence Index = 22}
	\label{Fig:05}
\end{figure}

Fig.\ref{Fig:04} and Fig.\ref{Fig:05} illustrate the PRACH performance in the presence of intra-cell interference (RootSequenceIndex equal to 22) for various Preamble Index  values. The objective behind these results is to examine how the PRACH's robustness is affected by the Preamble Index  of the Interfering UE (I-UE) (i.e., the UE generating the interference). It is important to note that all simulations in this section \label{sec:sss} were conducted with a Preamble Index of the Target UE (T-UE) set to 32.

Based on the results depicted in Fig.\ref{Fig:04} and Fig.\ref{Fig:05}, the following observations can be made:

\begin{description}
	\item [\textsl{(O1)}] \textsl{\textcolor{myColor}{Observation 1}}: NR-PRACH performance, specifically the preamble detection probability, decreases as the intra-cell interference level increases (i.e., when  SNRdB value of I-UE increases).
	\item [\textsl{(O2)}] \textsl{\textcolor{myColor}{Observation 2}}: Regardless of the specific Preamble Index used by the ``Interfering'' UE  (I-UE), NR-PRACH performance deteriorates as the intra-cell interference level increases.
\end{description}

\begin{figure}[htbp]
	\centering
	\fbox{
\includegraphics[page = 1,clip, trim=0.0cm 0.0cm 0.0cm 0.0cm, width=0.99\textwidth]{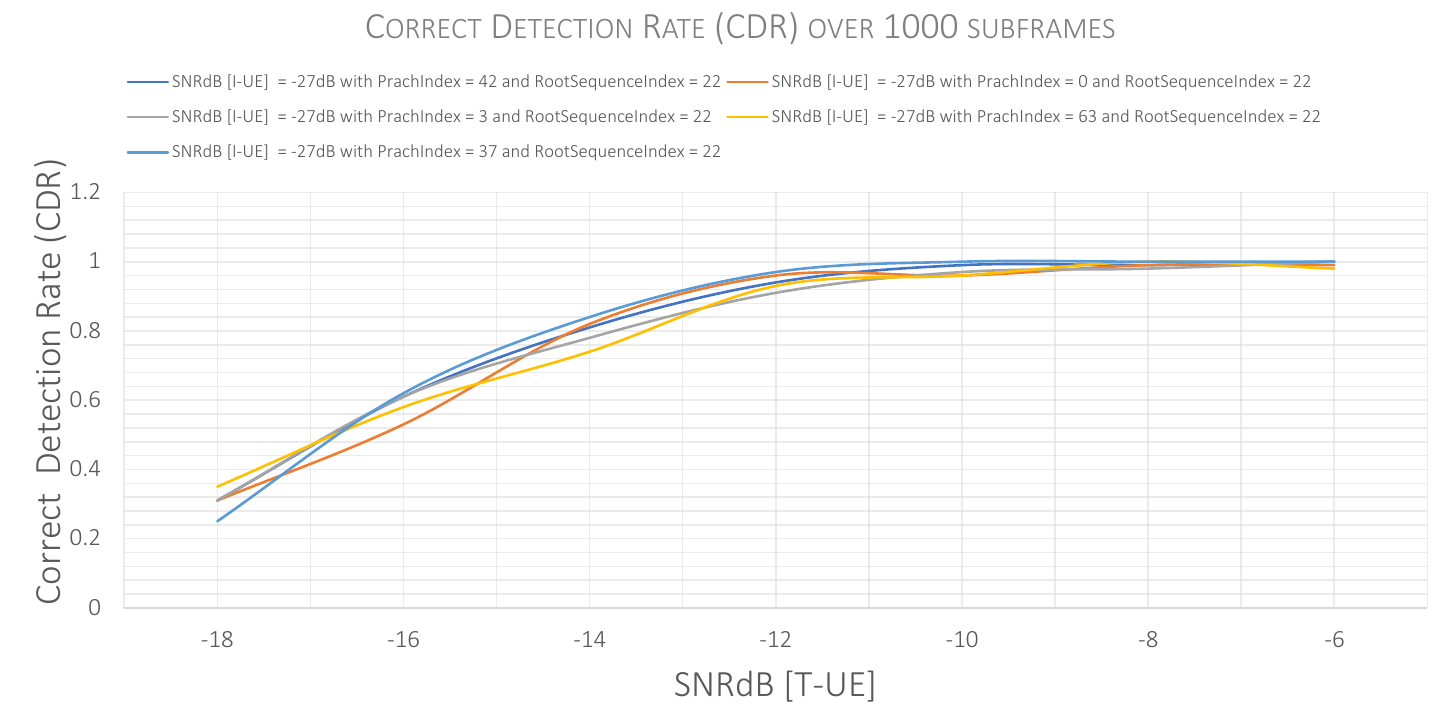}
}
	\caption{PRACH Performance in presence of Intra-Cell Interference with a Root Sequence Index of 22 and SNRdB [I-UE] of -27 dB for various ``Preamble Index'' values.}
	\label{Fig:06}
\end{figure}

\begin{figure}[htbp]
	\centering
	\fbox{
\includegraphics[page = 1,clip, trim=0.0cm 0.0cm 0.0cm 0.0cm, width=0.99\textwidth]{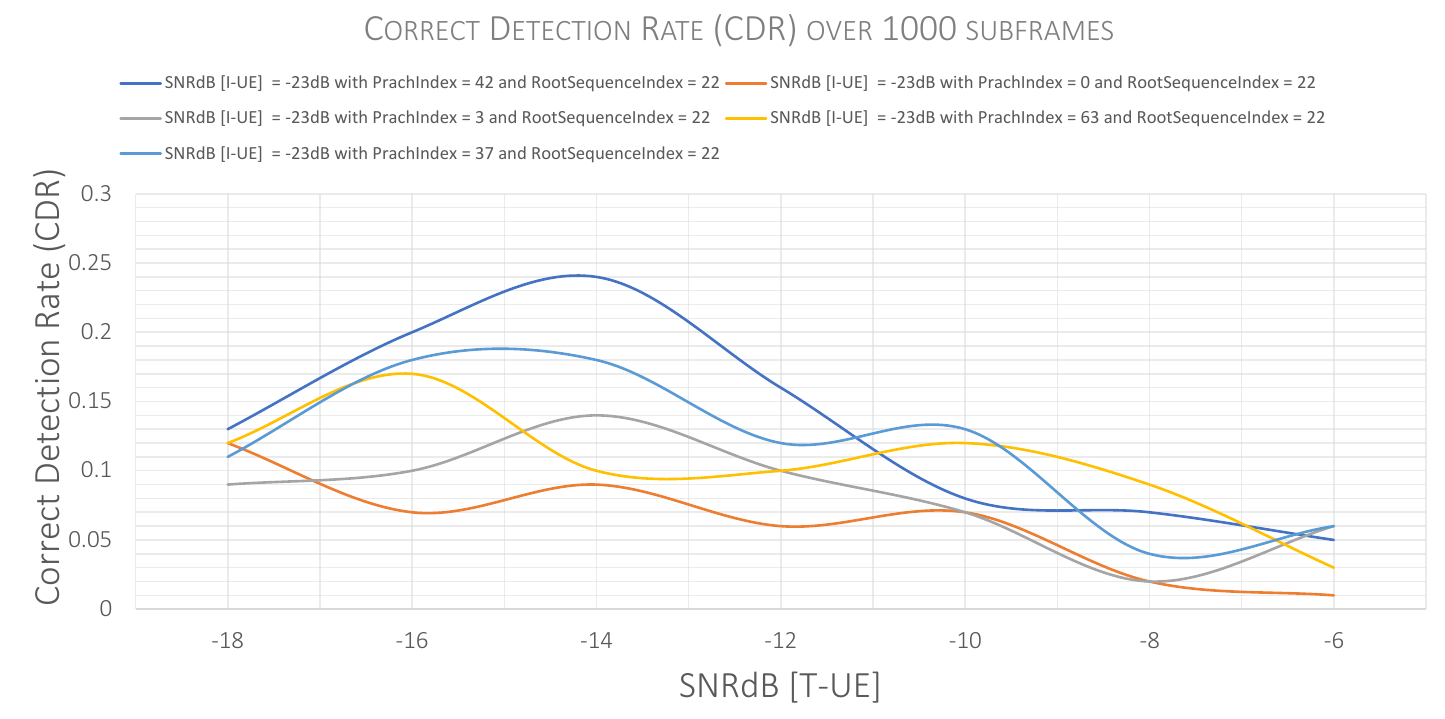}
}
	\caption{PRACH Performance in presence of Intra-Cell Interference with a Root Sequence Index of 22 and SNRdB [I-UE] of -23 dB for various ``Preamble Index'' values.}
	\label{Fig:07}
\end{figure}

Fig.\ref{Fig:06} and Fig.\ref{Fig:07} illustrate the performance of PRACH in the presence of intra-cell interference (RootSequenceIndex=22) for various ``Preamble Index'' values while maintaining constant the level of Intra-Cell Interference. The objective behind these results is to determine which ``Preamble Index'' values enable the ``Target'' UE (T-UE) to better withstand Intra-Cell Interference. Based on the findings from Fig.\ref{Fig:06} and Fig.\ref{Fig:07}, the following observations can be made:

\begin{description}
	\item [\textsl{(O3)}] \textsl{\textcolor{myColor}{Observation 3}}:At a relatively low intra-cell interference level (SNRdB [I-UE] = -27dB, as shown in Fig.\ref{Fig:06}), the performance of NR-PRACH remains constant regardless of the ``Preamble Index'' values utilized by the Interfering UE (I-UE).
	\item [\textsl{(O4)}] \textsl{\textcolor{myColor}{Observation 4}}: However, at a moderately elevated level of intra-cell interference (SNRdB [I-UE] = -23dB, as illustrated in Fig.\ref{Fig:07}), the performance of NR-PRACH becomes sensitive to the ``Preamble Index'' values employed by the ``Interfering UE'' (I-UE). 
\end{description}

\subsection{NR-PRACH Performance under Inter-Cell Interference Conditions}
\label{sec:sss2}
As was the case in  \ref{sec:sss1}, in this section, the performance of NR-PRACH is also evaluated through the Correct Detection Rate (CDR) over 1000 subframes. Inter-cell interference are simulated using a ``Interfering'' UE  (I-UE) located in a neighboring cell to whose of the ``Target'' UE  (T-UE), which means it uses a different value of the ``Logical Sequence Index'' and can potentially use the same ``Preamble index'' as the ``Target'' UE  (T-UE).

\begin{figure}[htbp]
	\centering
	\fbox{
\includegraphics[page = 1,clip, trim=0.0cm 0.0cm 0.0cm 0.0cm, width=0.99\textwidth]{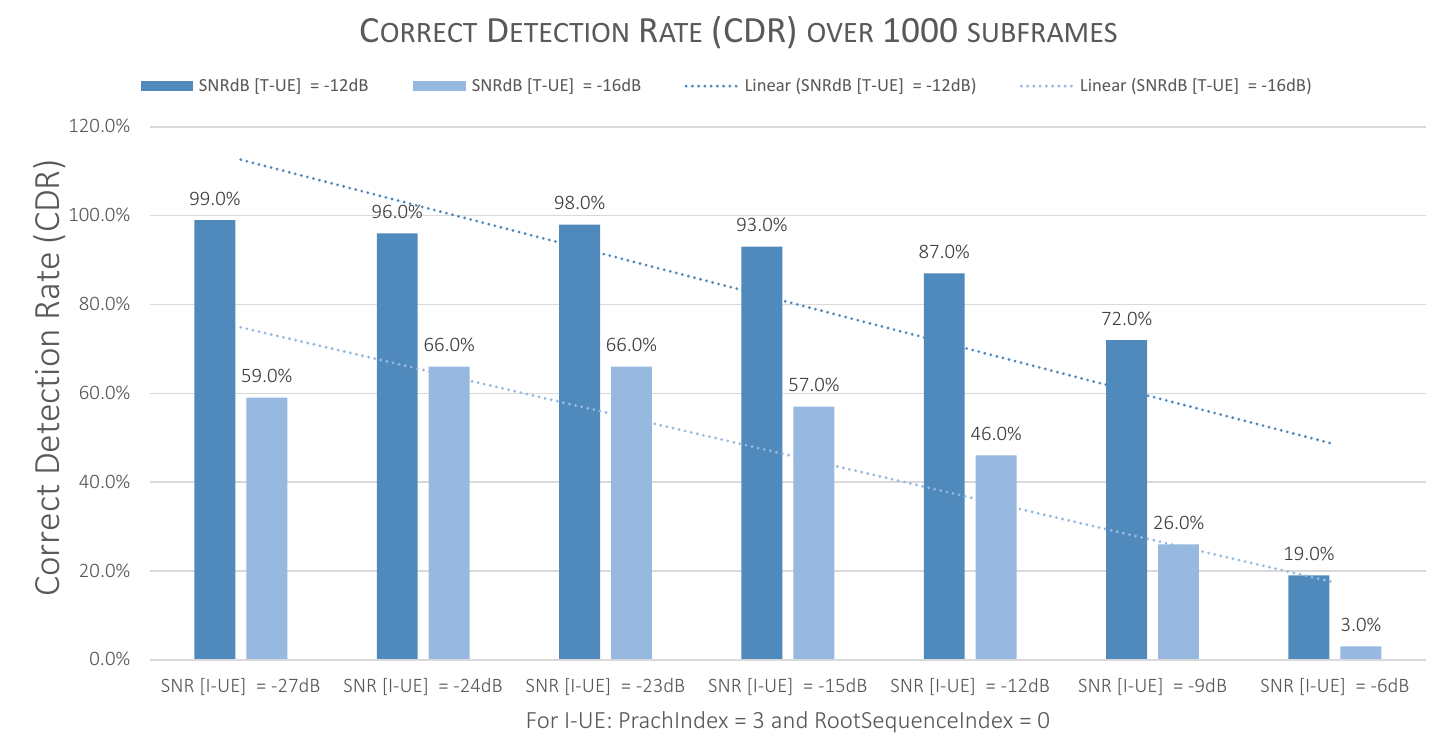}
}
	\caption{PRACH Performance in the Presence of Inter-Cell Interference with ``Preamble Index'' = 3 and Root Sequence Index = 0.}
	\label{Fig:08}
\end{figure}

\begin{figure}[htbp]
	\centering
	\fbox{
\includegraphics[page = 1,clip, trim=0.0cm 0.0cm 0.0cm 0.0cm, width=0.99\textwidth]{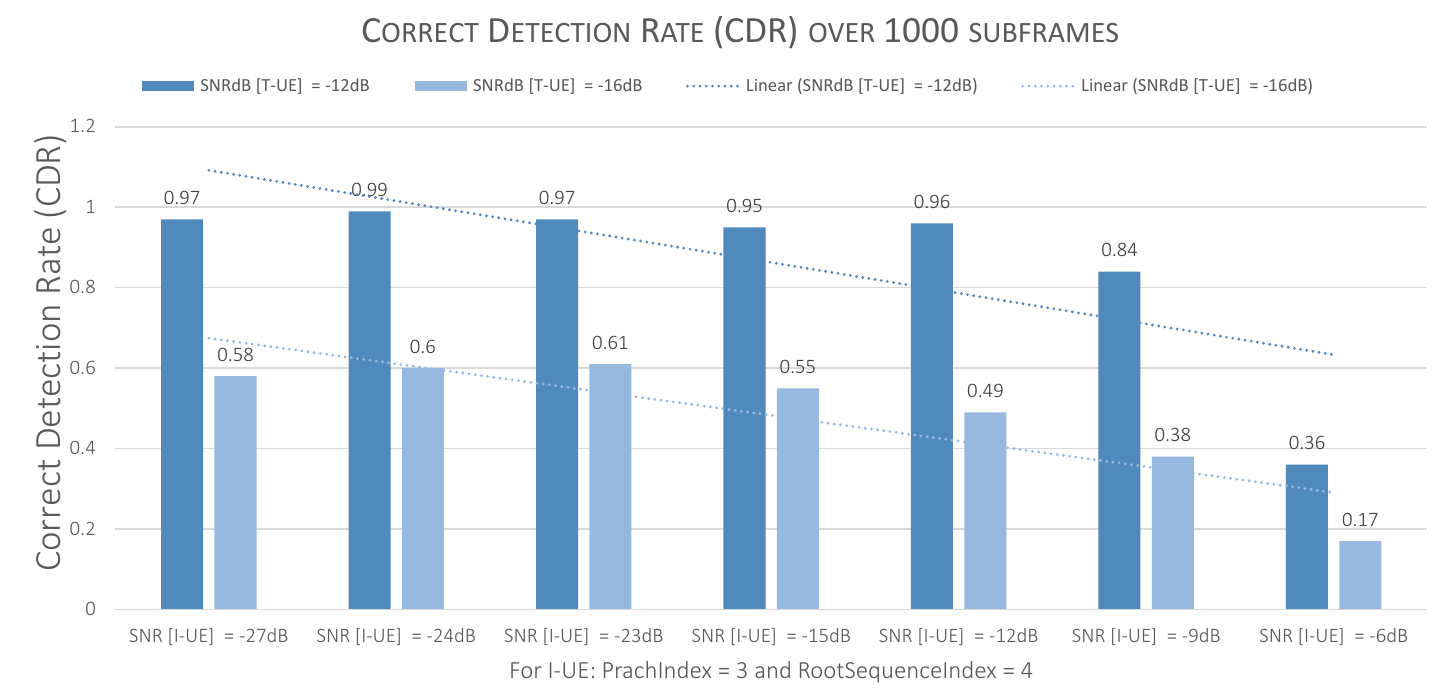}
}
	\caption{PRACH Performance in the Presence of Inter-Cell Interference with ``Preamble Index'' = 3 and Root Sequence Index = 4.}
	\label{Fig:09}
\end{figure}

Fig.\ref{Fig:08} and Fig.\ref{Fig:09}  illustrate the performance of PRACH in the presence of inter-cellr interference (RootSequenceIndex=22 for the``Target'' UE  (T-UE) and RootSequenceIndex = 0 and 4 for the ``Interfering'' UE  (I-UE)) for a ``Preamble Index'' value equal to 3. 

Based on the results depicted in these figures, the following observations can be made:

\begin{description}
	\item [\textsl{(O5)}] \textsl{\textcolor{myColor}{Observation 5}}:The performance of NR-PRACH, i.e., the probability of preamble detection, decreases as the level of inter-cellinterference increases (i.e., as the value of SNRdB [I-UE] increases).
	\item [\textsl{(O6)}] \textsl{\textcolor{myColor}{Observation 6}}: Regardless of the RootSequenceIndex value used by the ``Interfering'' UE  (I-UE), the performance of NR-PRACH decreases as the level of inter-cell interference increases. This observation aligns with Observation 2. 
\end{description}

\begin{figure}[htbp]
	\centering
	\fbox{
\includegraphics[page = 1,clip, trim=0.0cm 0.0cm 0.0cm 0.0cm, width=0.99\textwidth]{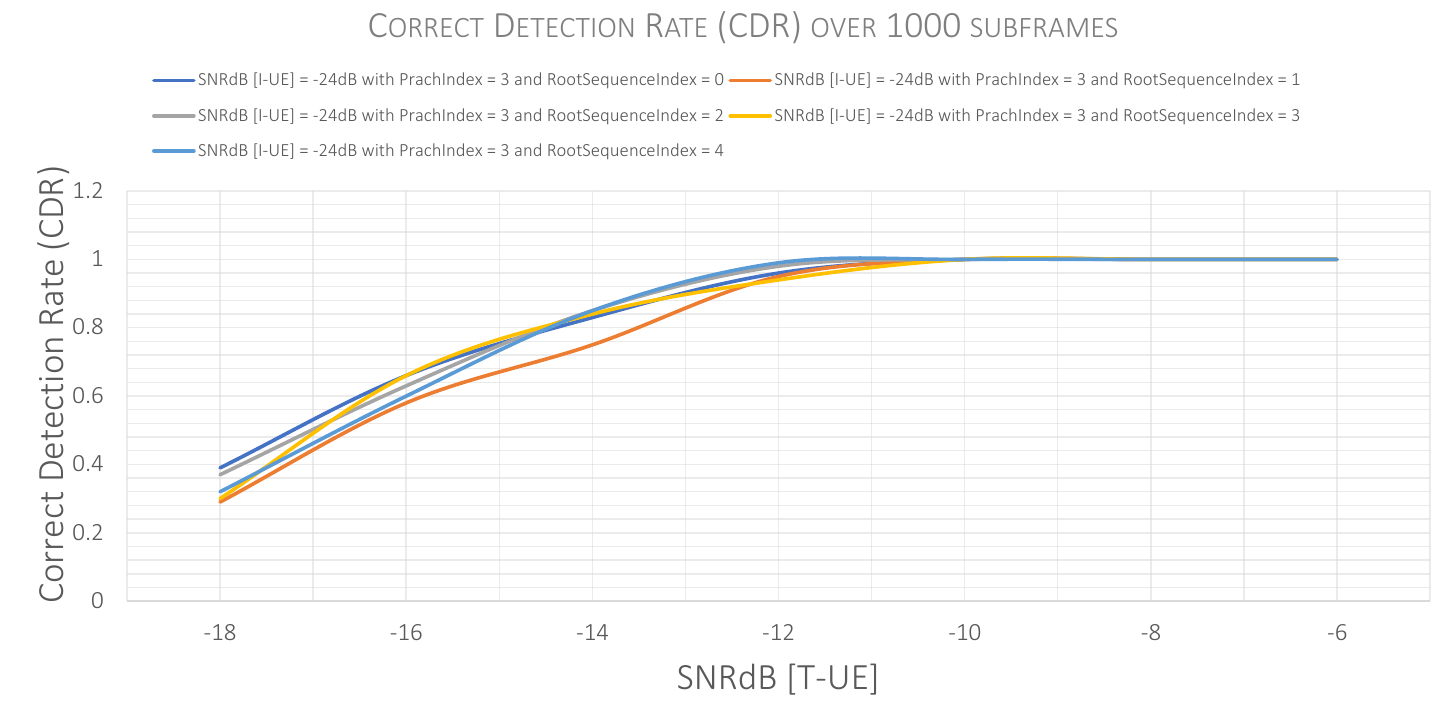}
}
	\caption{PRACH Performance in the Presence of Inter-Cell Interference with ``Preamble Index'' = 3 and SNRdB [I-UE] = -24dB for different Values of RootSequenceIndex for the Interfering Cell.}
	\label{Fig:10}
\end{figure}

\begin{figure}[htbp]
	\centering
	\fbox{
\includegraphics[page = 1,clip, trim=0.0cm 0.0cm 0.0cm 0.0cm, width=0.99\textwidth]{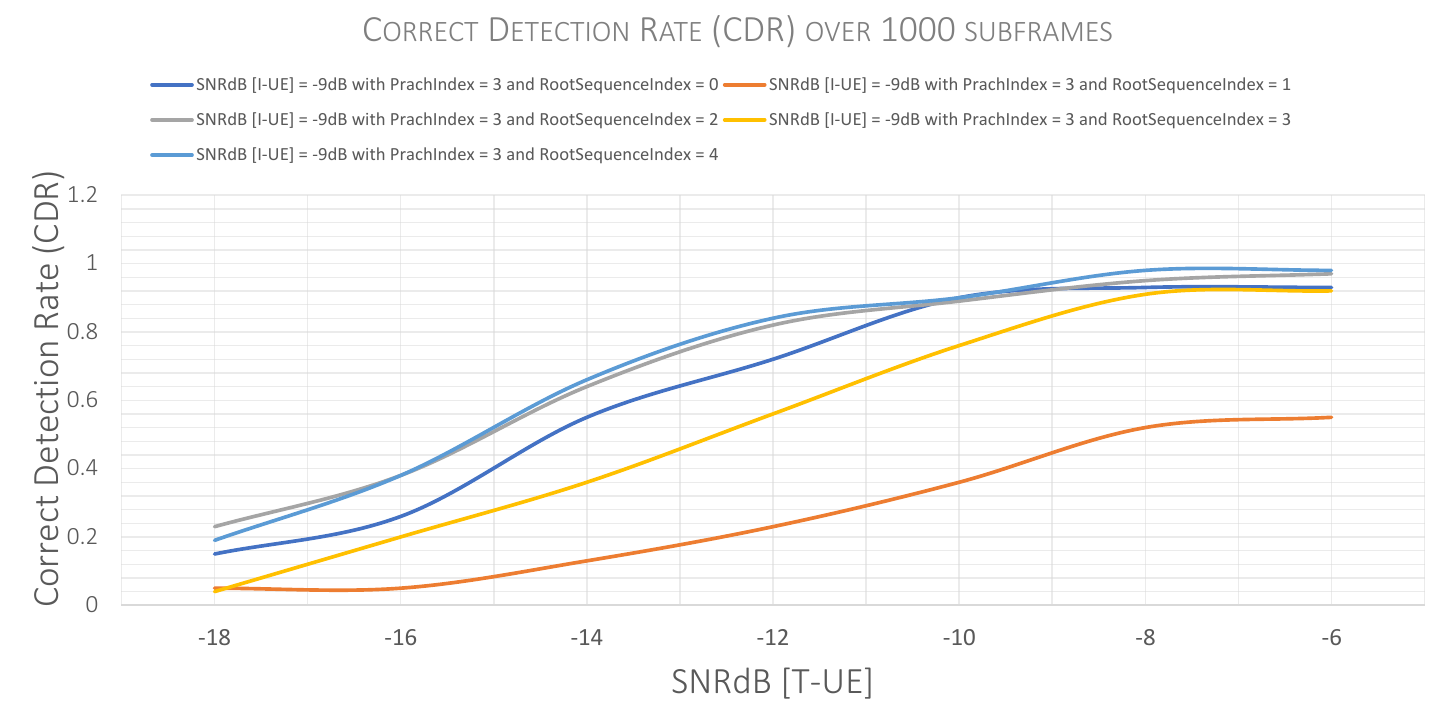}
}
	\caption{PRACH Performance in the Presence of Inter-Cell Interference with ``Preamble Index'' = 3 and SNRdB [I-UE] = -9dB for different Values of RootSequenceIndex for the Interfering Cell.}
	\label{Fig:11}
\end{figure}

Fig.\ref{Fig:10} and Fig.\ref{Fig:11} depict the performance of PRACH in the presence of inter-cell interference (RootSequenceIndex=22 for the ``Target'' UE  (T-UE)) for various values of RootSequenceIndex for the ``Interfering'' UE  (I-UE) but at a constant level of inter-cell interference. The objective of these results is to examine which RootSequenceIndex values enable the T-UE to better withstand intra-cell interference.

Based on the findings from these figures, the following observations can be made:

\begin{description}
	\item [\textsl{(O7)}] \textsl{\textcolor{myColor}{Observation 7}}: At a relatively low level of intra-cell interference (SNRdB [I-UE] = -24dB, refer to Fig.\ref{Fig:10}), the performance of NR-PRACH remains unchanged regardless of the ``RootSequenceIndex'' values used by the ``Interfering'' UE  (I-UE).
	\item [\textsl{(O8)}] \textsl{\textcolor{myColor}{Observation 8}}: However, at a moderately high level of intra-cell interference (SNRdB [I-UE] = -9dB, refer to Fig.\ref{Fig:11}), the performance of NR-PRACH becomes sensitive to the ``RootSequenceIndex'' values employed by the interfering mobile. For instance, the results in Figure 34 demonstrate that a ``RootSequenceIndex'' of 1 significantly degrades the performance of NR-PRACH compared to a ``RootSequenceIndex'' of 4. 
\end{description}

%
%
%
%

\subsection{Outcomes and  Future of PRACH Detection}
	\label{sec:sss3}
The observations provided in \ref{sec:sss1} and \ref{sec:sss2} suggest that both intra-cell and inter-cell interference levels can significantly affect the performance of NR-PRACH in a wireless communication system. The choice of certain parameters by the ``Interfering'' UE  (I-UE), such as RootSequenceIndex or ``Preamble Index'', can exacerbate or mitigate these effects, depending on the specific interference scenario. Understanding and managing interference levels and parameters selection are crucial for optimizing the performance of NR-PRACH in practical deployments.

AI/ML hold immense promise for the future of PRACH detection and management \cite{RFang2022}. These technologies can enhance detection accuracy, improve interference management, optimize resource allocation, and enable PRACH systems to adapt dynamically to the complexities of modern wireless networks. The integration of AI/ML will be crucial in achieving efficient and reliable PRACH operations in the rapidly evolving world of wireless communications.

\section{Conclusion}


This article explores the impact of UE (User Equipment) and cell configuration parameters on interference. Observation 1 and Observation 2 emphasize that NR-PRACH performance, specifically the preamble detection probability, deteriorates as intra-cell interference levels increase. This underscores the importance of managing interference within the cell to maintain reliable communication. Observation 5 and Observation 6 extend this concern to inter-cell interference, indicating that NR-PRACH performance also declines as inter-cellular interference levels rise. This emphasizes the need for effective interference management strategies not only within cells but also between neighboring cells in 5G NR networks.

These findings underscore the critical role of interference management and the configuration of UE/Cell parameters in ensuring the robustness and reliability of the NR-PRACH performance in 5G NR networks. Optimizing these configurations can lead to improved network performance and enhanced communication quality (QoS).

\section{Perspectives}

In the rapidly evolving landscape of 5G wireless communication systems, addressing the challenges posed by intra and inter-cell interference in 5G-NR PRACH  detection performance is paramount to ensure the seamless and efficient operation of future networks. In terms of future works, the following  ideas might be considered.
\begin{description}
	\item [(1)] \textsl{\textcolor{myColor}{Adaptive Preamble Index Assignment:}} Develop an adaptive algorithm that dynamically assigns preamble indexes to UEs based on the current interference level, aiming to maximize NR-PRACH performance. This algorithm could take into account both intra-cell and inter-cell interference.
	\item [(2)] \textsl{\textcolor{myColor}{Machine Learning-Based Interference Mitigation:}} Investigate the use of machine learning techniques, such as deep learning or reinforcement learning, to predict and mitigate interference in NR-PRACH. Train models on historical interference data and evaluate their effectiveness.
	\item [(3)] \textsl{\textcolor{myColor}{Dynamic Power Control:}} Propose a dynamic power control mechanism that adjusts the transmission power of UEs based on real-time interference measurements. This could help manage interference levels and improve NR-PRACH performance.
	\item [(4)] \textsl{\textcolor{myColor}{Cooperative Interference Management:}} Explore cooperative interference management techniques where neighboring cells collaborate to mitigate interference for NR-PRACH. This could involve information sharing and joint decision-making among base stations.
	\item [(5)] \textsl{\textcolor{myColor}{Frequency Domain Interference Mitigation:}} Research methods that leverage frequency domain processing to separate and mitigate interference signals, especially in scenarios with overlapping frequency bands.
	\item [(6)] \textsl{\textcolor{myColor}{Cognitive Radio Approaches:}} Explore cognitive radio-based approaches that enable UEs to dynamically select PRACH resources and parameters based on the interference environment. Implement and evaluate cognitive PRACH algorithms.
	\item [(7)] \textsl{\textcolor{myColor}{Machine Learning for Interference Prediction:}} Develop machine learning models that predict interference patterns in real-time based on historical data and environmental factors. Use these predictions to optimize NR-PRACH configurations.
\end{description}

\newpage

\bibliographystyle{vancouver}

\end{document}